\documentstyle[11pt,newpasp,twoside,epsf]{article}
\def\edcomment#1{\iffalse\marginpar{\raggedright\sl#1\/}\else\relax\fi}
\reversemarginpar

\begin{document}
\title{The Mass Distribution in Low Surface Brightness Galaxies}
\author{R.A. Swaters}
\affil{Carnegie Institution of Washington, 5241 Broad Branch Rd. NW,
  Washington DC 20015, U.S.A.; swaters@dtm.ciw.edu}

\begin{abstract}
  The distribution of the stellar and gaseous components in low
  surface brightness galaxies has been determined directly from
  optical and HI imaging. The distribution of what might be the
  dominant mass component, the dark matter, which is inferred from
  rotation curves, is far harder to determine. Although the rotation
  curves themselves can be determined fairly accurately from HI
  synthesis observations, and in particular from H$\alpha$
  spectroscopy, the uncertainty in the mass modeling leaves room for
  a wide range of possible dark matter distributions, ranging from
  maximum stellar disks with shallow dark halos to cuspy dark halos
  with little mass in stars.
\end{abstract}

\section{Introduction}

Low surface brightness (LSB) galaxies have been receiving an
increasing amount of attention since their discovery, in part because
their properties substantially increase the range in galaxy properties
over which theories of galaxy structure, formation and evolution can
be tested.  In particular, the rotation curves of these galaxies
provide important clues to the distribution of dark matter.

\section{The visible mass distribution}

The visible mass distribution in LSB galaxies can easily be measured
directly by optical and HI imaging. Optical surface photometry (e.g.,
McGaugh \& Bothun 1994, de Blok etal.\ 1995) has shown that the
surface brightness of LSB galaxies is well below the average value of
21.65 $B$-mag arcsec$^{-2}$ for high surface brightness (HSB) galaxies
(Freeman 1970). The shape of the light distribution is remarkably
similar between LSB and HSB galaxies. Both types of galaxies are
dominated by exponential disks, and both may have significant bulges
(Beijersbergen et al.\ 1999). The disk scale lengths found in LSB
galaxies are in the same range as those for HSB galaxies, indicating
that LSB galaxies are not dwarf galaxies. At a fixed luminosity,
however, LSB galaxies have larger scale lengths than HSB galaxies.

The HI distribution for a sample of LSB galaxies has been mapped by de
Blok et al.\ (1996). In HI, as in the optical, the shapes of the HI
distribution seen in LSB galaxies is similar to those seen in HSB
galaxies, ranging from filled HI disks, to disks with central holes in
HI or ring-like HI distributions. A key difference between HI in LSB
and HSB galaxies is that LSB galaxies on average have HI densities
that are about a factor of two lower than their HSB counterparts. The
HI column densities thus appear related to the surface brightness (see
also Swaters et al. 2000a).

\section{The total mass distribution}

A first estimate of the total mass density has been obtained from the
Tully-Fisher relationship for LSB galaxies. Zwaan et al.\ (1995) found
that LSB galaxies follow the same Tully-Fisher relationship as HSB
galaxies, and concluded from this that LSB galaxies must have lower
total mass densities than HSB galaxies.

This result was placed on firmer footing when rotation curves for LSB
galaxies became available. De Blok et al.\ (1996) and de Blok \&
McGaugh (1996), on the basis of HI observations, found that the
rotation curves of LSB galaxies rise more slowly that those of HSB
galaxies if the radii are expressed in units of kpc, confirming the
Zwaan et al.\ (1995) result. LSB galaxies therefore appear to be true
low density galaxies.

\section{The dark mass distribution}

Although it is clear that LSB galaxies have lower surface
brightnesses, lower HI densities and lower total densities than HSB
galaxies, this does not automatically imply that these galaxies also
have lower dark matter densities. Detailed mass modelling is required
to obtain limits on the dark matter distribution.

De Blok \& McGaugh (1997) derived mass models for LSB galaxies from HI
rotation curves, and found that LSB galaxies have to be dominated by
dark matter, both because the HI rotation curve shapes ruled out
maximum disks, and because the inferred stellar mass-to-light ratios
($\Upsilon$) seem inconsistent with other indicators of $\Upsilon$ in
these galaxies.

More recent high resolution observations of the rotation curves of LSB
galaxies (Swaters et al.\ 2000b) and a reanalysis of the original HI
observations (van den Bosch et al. 2000) have indicated that the HI
rotation curves were affected by beam smearing. Even though the
H$\alpha$ rotation curves presented in Swaters et al.\ (2000) rise
somewhat more steeply than the HI rotation curves of the same
galaxies, they do confirm the main conclusion by de Blok et al.\ 
(1996) that LSB galaxies have low total mass densities. However, the
new H$\alpha$ rotation curves do change the conclusions derived from
the mass modelling.

\begin{figure}[t]
\plotone{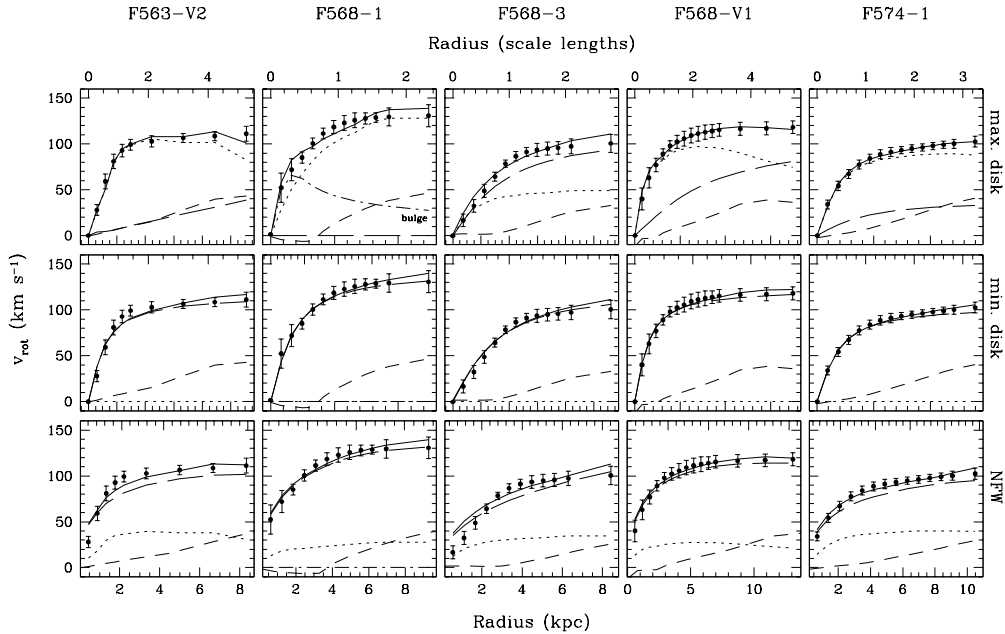}
\caption{{\small Mass models for 5 LSB galaxies. From top to bottom the
  maximum disk and minimum disk mass models with isothermal halos, and
  a mass model with an NFW halo and $\Upsilon=1.0$. The dotted line
  represent the contribution of the stellar disk, the short dashed
  line that of the gas, the long dashed line the dark matter, and the
  full line gives the best fit mass model. }}
\vskip-\baselineskip
\end{figure}

In Fig.~1 three mass models are shown for the five galaxies for which
Swaters et al.\ (2000) presented H$\alpha$ rotation curves.
Interestingly, in all galaxies except F568-3, the contribution of the
stellar disk can be scaled to explain the inner parts of the rotation
curve. The derived values for $\Upsilon$ may be high: up to 17 in the
$R$-band. Most of these are outside the range of what current
population synthesis models predict.  If these high values of
$\Upsilon$ are to be explained solely by a stellar population, the
stellar content and the processes of star formation in LSB galaxies
need to be very different from those in HSB galaxies.  Alternatively,
these high mass-to-light ratios might indicate the presence of an
additional baryonic component associated with the disk, suggested to
be in the form of cold molecular gas (e.g., Pfenniger et al.\ 1994) or
white dwarfs (e.g., Ibata et al.\ 1999).  On the other hand, the fact
that the stellar disk can be scaled to explain the observed rotation
curve may simply reflect the possibility that the luminous and dark
mass have a similar distribution within the optical galaxy.

The other extreme for the contribution of the stellar disk to the
rotation curve is to assume that its contribution is negligible. From
Fig.~1 it is clear that the minimum disk mass models fit the rotation
curves equally well as the maximum disk models.  In fact, good fits
can be obtained with any mass-to-light ratio lower than the maximum
disk one, demonstrating that the degeneracy that exists in the mass
modeling for HSB galaxies also exists for LSB galaxies.

If the contribution of the stellar disk is close to maximum, then the
derived core radii for the dark halos are large, and the central
densities low. In the minimum disk case, on the other hand, high
central dark matter densities, similar to those found in HSB galaxies,
and small core radii are required to explain the observed rotation
curves.  Because of this large degeneracy it is not possible to
determine the distribution of dark matter from rotation curves alone,
nor is it possible to determine whether the dark matter density in LSB
galaxies scales with luminous or total mass density. If the fractional
contribution of the stellar disk to the rotation curve is similar in
all galaxies, then the central dark matter density does scale with
surface brightness. If, on the other hand, HSB galaxies are close to
maximum disk, and LSB close to minimum disk, which is consistent with
the values for $\Upsilon$ expected from current stellar population
synthesis models, then LSB and HSB galaxies may well have identical
halos, and hence the dark matter density is not correlated with
surface brightness.

Another possible model for the dark matter distribution is shown in
the bottom row of Fig.~1. Here, mass models are displayed that are
based on halo profiles with a radial distribution as suggested by
Navarro et al.\ (1997), which have a steep slope of $r^{-1}$ in the
inner parts.  This profile has been derived from CDM simulations, and
it has been successful in explaining the rotation curves of HSB
galaxies.  The HI data indicated that these NFW halos are inconsistent
with the observed rotation curves (e.g., McGaugh \& de Blok 1998),
which has been one of the reasons for the recent surge in finding
alternatives for CDM.

The NFW mass models in Fig.~1 have been derived for $\Upsilon=1$, and
have been corrected for adiabatic contraction. These NFW mass models
fit the observed rotation curves nearly as well as those based on
isothermal halos, except for F568-3. The derived concentration
parameters $c$ for the NFW fits are consistent with a $\Lambda$CDM
cosmology ($9<c<16$). This demonstrates that LSB galaxies may have
cuspy halos, despite the fact that they appear to be low density
objects. There is still controversy on this point. De Blok et
al.\ (this conference) find that a substantial fraction of their new
H$\alpha$ observations are inconsistent with an NFW halo. Pickering
et al.\ (this conference) on the other hand, find that most of the
H$\alpha$ rotation curves for their sample of about 70 LSB galaxies
can be fitted with a cosmologically meaningful NFW profile. At this
point it is unclear whether this apparent dichotomy is the result of
observational uncertainties or non-circular motions in the inner parts
of the galaxies, or whether the rotation curves of LSB galaxies
are in fact inconsistent with cuspy halos.

The fact that the three mass models discussed above, as well as many
others, can provide good fits to the observed rotation curves of LSB
galaxies, reflects the large degree of freedom that exists in the mass
modeling. Based on rotation curves alone, even with the high spatial
resolution of H$\alpha$ observations, it is difficult to get useful
constaints on the dark matter properties. Independent estimates of the
mass in the stellar disks, obtained for example from stellar velocity
dispersions, are essential.

\vskip-2\baselineskip

\end{document}